\documentclass[aps,preprint,amsmath,amssymb,amsfonts,showpacs]{revtex4}
\usepackage{amsthm}
\usepackage{epsfig}
\usepackage{graphicx}
\usepackage{subfigure}
\usepackage{dcolumn}
\usepackage{bm}

\newcommand{\del}{\partial}

\begin{document}

\title{A look at area Regge calculus}

\author{Yasha Neiman}
\email{yashula@gmail.com}
\affiliation{Institute for Gravitation \& the Cosmos and Physics Department, Penn State, University Park, PA 16802, USA}

\date{\today}

\begin{abstract}
Area Regge calculus is a candidate theory of simplicial gravity, based on the Regge action with triangle areas as the dynamical variables. It is characterized by metric discontinuities and vanishing deficit angles. Area Regge calculus arises in the large-spin limit of the Barrett-Crane spinfoam model, but not in the newer EPRL/FK model. We address the viability of area Regge calculus as a discretization of General Relativity. We argue that when all triangles are spacelike and all tetrahedra have the same signature, non-trivial solutions of the area calculus are associated with a nonzero Ricci scalar. Our argument rests on a seemingly natural regularization of the metric discontinuities. It rules out the Euclidean area calculus, as well as the Lorentzian sector with all tetrahedra spacelike - the two setups usually considered in spinfoam models. On the other hand, we argue that the area calculus has attractive properties from the point of view of finite-region observables in quantum gravity. 
\end{abstract}


\maketitle

\section{Introduction} \label{sec:intro}

\begin{figure}%
\centering%
\includegraphics[scale=0.75]{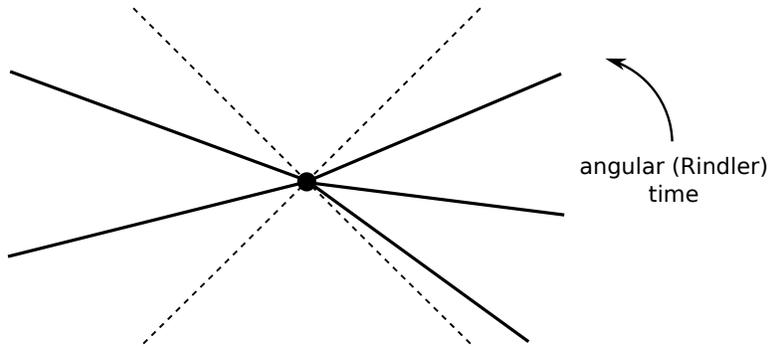} \\
\caption{The 1+1d plane normal to a spacelike triangle (central dot) in a Lorentzian simplicial manifold. The triangle is shared by several tetrahedra (solid lines) and an equal number of 4-simplices (wedges between the solid lines). The two dashed lines depict the two null normals to the triangle. In this example, the tetrahedra sharing the triangle are all spacelike.}
\label{fig:triangle} 
\end{figure}

In classical General Relativity (GR), discretized spacetimes are a useful approximation tool. Regge calculus \cite{Regge:1961px} is an elegant formulation of discretized GR. In it, spacetime is approximated with a simplicial decomposition. The combinatorial structure of the triangulation is held fixed, with the edge lengths as the dynamical variables. The edge lengths function as discrete analogues of metric components. Each 4-simplex separately is flat, its geometry fixed by its 10 edge lengths. Curvature is encoded in the deficit angles around triangular 2-faces. These are constructed from the dihedral angles between tetrahedra within the 4-simplices surrounding each triangle. See figure \ref{fig:triangle}. The bulk action takes the form:
\begin{align}
 S_{\text{Regge}} = \sum_f A_f \bigg(\Theta_f^{(0)} - \sum_{v\in f}\theta_{vf} \bigg) \ .
\end{align}
Here, $f$ denotes the triangles (faces of the dual complex), and $v$ denotes the 4-simplices (vertices of the dual complex). The $A_f$ are the triangles' areas. $\theta_{vf}$ is the dihedral angle between the tetrahedra sharing the triangle $f$ in the 4-simplex $v$. $\Theta_f^{(0)}$ is the flat value of the sum $\sum\theta_{vf}$ around the triangle $f$. It depends on the signature of the triangle's normal plane, as well as on conventions, if the plane is Lorentzian. We work in units where $c = \hbar = 8\pi G = 1$. 

Regge calculus is closely related to a number of attempts at quantizing gravity \cite{Regge:2000wu}. In particular, it has ties with the loop quantum gravity (LQG) \cite{Rovelli:2004tv,Thiemann:2007zz} and spinfoam \cite{Perez:2012wv} approaches, which have merged in the EPRL/FK spinfoam model \cite{EPRL,FK}. In LQG, spatial hypersurfaces are naturally discretized into quantum analogues of polyhedra \cite{Bianchi:2010gc}. The areas of the 2-faces are determined by spin labels. In addition, each polyhedron carries an intertwiner, which determines (up to quantum uncertainty) the dihedral angles between the 2-faces. In spinfoam models, one discretizes space\emph{time}, again with spin and intertwiner quantum numbers on the 2-faces and 3-faces. The spins and intertwiners in the bulk of the spacetime region are summed over. While it is clear that one should also sum over the combinatorics of the discretization, there is no agreement about which combinatorics should be allowed. A simple possibility is to demand that the spinfoams correspond to simplicial decompositions. This establishes a direct link to Regge calculus.

An obvious difference between Regge calculus and LQG/spinfoams is that in the latter, there is no concept of edge lengths, or indeed of edges. The combinatorics and the quantum numbers all refer to 2-faces, 3-faces and 4-volumes. Early on, Carlo Rovelli suggested \cite{Rovelli:1993kc} that LQG is related to a new type of Regge calculus, now known as area Regge calculus. In area Regge calculus, the fundamental variables are not the edge lengths, but the triangle areas. Since a 4-simplex contains both ten edges and ten triangles, the areas can determine its geometry just as well as the edge lengths (at least on the level of degree-of-freedom counting; the correspondence is not one-to-one when right angles are involved \cite{Barrett:1997tx}). 

It turns out \cite{Barrett:1997tx} that area Regge calculus is fundamentally different from the standard length-based calculus. First, the deficit angles are forced to vanish on-shell. Second, the theory allows for metric discontinuities. This is in contrast to other variants of Regge calculus (e.g. \cite{Barrett:1994nn,Dittrich:2008va}), which simply rewrite the same dynamics in different variables. The area calculus is \emph{also} different from canonical LQG and from the EPRL spinfoam. While LQG spins are directly analogous to the triangle areas, there is no variable in the area calculus that would correspond to LQG intertwiners. Nevertheless, there exists a quantum gravity model to which area Regge calculus is relevant. That model is the Barrett-Crane spinfoam \cite{Barrett:1997gw,Barrett:1999qw}, the predecessor of EPRL. We review its relation to area Regge calculus and to the EPRL model in section \ref{sec:review:BC}.

The question whether area Regge calculus is a viable discretization of GR has so far remained open. In section \ref{sec:review:AreaRegge}, we review the state of the art on this issue. We then proceed to expand on it in two ways. In section \ref{sec:concept}, we discuss the conceptual advantages of area Regge calculus in discussing finite-region observables in quantum gravity. We argue that it allows notions of evolution of 3d hypersurfaces, and especially of 2d surfaces, that avoid the ``problem of time'' associated with transverse diff-invariance. 

Despite this, in section \ref{sec:Ricci}, we indicate an apparent technical failure of the theory. After establishing a seemingly natural regularization scheme, we argue that non-trivial solutions of Euclidean area Regge calculus always carry nonzero scalar curvature $R\neq 0$ at the triangles. This scalar curvature has the same sign at each triangle, and thus cannot be averaged away. This is in contradiction to pure GR, whose solutions have $R=0$. The argument also extends to Lorentzian spacetime, if all the triangles are spacelike and all the tetrahedra have the same signature. This includes the case of spacelike tetrahedra, which is the one usually considered in Lorentzian spinfoams. One could hope that a nonzero Ricci scalar is simply an indicator of a cosmological constant. However, this possibility is also problematic. We discuss it along with several others in section \ref{sec:discuss}. It appears, then, that area Regge calculus is not a valid discretization of GR, at least not in the sectors of interest for either Euclidean or Lorentzian spinfoams. We thus end up agreeing with the abandonment of the Barrett-Crane model.

\section{Background} \label{sec:review}

\subsection{Area Regge calculus} \label{sec:review:AreaRegge}

Within a single 4-simplex, standard Regge calculus and area Regge calculus agree. However, they lead to quite different pictures on a larger triangulation \cite{Barrett:1997tx}. In the area calculus, two neighboring 4-simplices may disagree on the geometry of their shared tetrahedron. This is because a tetrahedron's geometry has six degrees of freedom, out of which its face areas supply only four. Thus, in the area calculus, the metric is discontinuous across 4-simplex boundaries. On the other hand, the equations of motion of the area calculus force all the deficit angles to vanish:
\begin{align}
 \Theta_f^{(0)} - \sum_{v\in f}\theta_{vf} = 0 \ .
\end{align}
Thus, if it weren't for the discontinuities, we would say that all the solutions are flat. However, the discontinuities are important, and may provide an alternative source of curvature. 

Let us look closer at the nature of the metric discontinuities. Consider a triangle, shared by several tetrahedra and 4-simplices (figure \ref{fig:triangle}). The 4-simplices all agree on the triangle's area, but in general disagree on its angles. Thus, as we cross successive tetrahedra in the triangle's normal plane, the triangle's metric undergoes area-preserving deformations. Infinitesimally, such deformations are encoded by a $2\times 2$ traceless symmetric matrix $\sigma_{ij}$, i.e. by a shear tensor. These are the same degrees of freedom that encode the amplitude and polarization of a gravitational wave. 

Another peculiarity of the area calculus is that spatial hypersurfaces, considered in isolation from the bulk spacetime, do not have a determinate metric. In simplicial gravity, a spatial hypersurface is a triangulated 3d manifold composed of tetrahedra. Now, in standard Regge calculus, the geometry of these tetrahedra is determined by their edge lengths. In contrast, in the area calculus, one only has the face areas. As noted above, these do not suffice to determine a tetrahedron's geometry. The tetrahedra in the hypersurface only acquire a metric once we are given the areas of the other triangles in the 4-simplices that contain them. The metric of the hypersurface is then obtained by patching the tetrahedra together. Whenever neighboring tetrahedra in the hypersurface do not share a 4-simplex, the metric may be discontinuous across their boundary. Also, as with a single tetrahedron, the 4-simplices on either side of the hypersurface induce two different geometries. Thus, the hypersurface has two metrics: an ``initial'' one and a ``final'' one.

It has long been unclear whether area Regge calculus is a discretization of GR, or indeed of any continuum theory. The literature contains several positive indications. A perturbative analysis on a hypercubic lattice \cite{Regge:2000wu} shows that the number of true degrees of freedom is the same in standard Regge calculus and in the area calculus. An analysis of a simple class of solutions with a discontinuity across a single hyperplane \cite{Wainwright:2004yn} recovers the ``refractive wave'' geometries, introduced in \cite{Barrett:2000jy} as generalized solutions of GR. Moreover, the allowed discontinuities were found to agree with the junction conditions in linearized GR.

\subsection{The Barrett-Crane spinfoam and its large-spin limit} \label{sec:review:BC}

The Barrett-Crane spinfoam model uses spin degrees of freedom to encode triangle areas in a spacetime triangulation. Unlike the EPRL model, it has no intertwiner degrees of freedom to encode the angles between triangles in a tetrahedron. The large-spin limit of the Barrett-Crane 4-simplex amplitude \cite{Baez:2002rx,Freidel:2002mj,Barrett:2002ur} reduces to the action of area Regge calculus, with an important caveat: each 4-simplex separately can have the sign of its action flipped, or worse yet, it can fall into a degenerate configuration. In fact, the large-area limit of the amplitude is dominated by these degenerate contributions. 

It was shown in \cite{Bianchi:2006uf} that for simple triangulations, the degenerate contributions can be damped by a boundary state peaked both in areas and in dihedral angles. However, it's unclear in this approach how to avoid degenerate 4-simplices in the bulk of a large triangulation. Non-degenerate contributions with the opposite sign of the Regge action are also puzzling. See \cite{Rovelli:2012yy,Christodoulou:2012sm} for a discussion. A version of the model without sign-flipped actions was proposed in \cite{Livine:2002rh}. The sign-flipped action and the degenerate 4-simplices persist in the EPRL spinfoam \cite{Barrett:2009mw,Han:2013gna,Han:2013hna}. There exist proposals to overcome them by modifying the model \cite{Mikovic:2011zx,Engle:2011un,Rovelli:2012yy}. 

In this paper, we give the Barrett-Crane model the benefit of the doubt with regard to the ``wrong'' contributions to the large-spin limit and their implications. We focus on the ``correct'' contribution, which corresponds to area Regge calculus, and ask whether it is viable. This question is largely answered in the negative in the LQG community, in part due to the mismatch between the degrees of freedom in the Barrett-Crane model and the kinematical state space of LQG. This mismatch is exemplified in the 4-simplex ``graviton propagator'' calculation \cite{Alesci:2007tx}, which showed that the Barrett-Crane vertex doesn't respond correctly to the intertwiner data in the boundary state. On this issue, we again give the Barrett-Crane model the benefit of the doubt; we do not owe allegiance to the LQG kinematical Hilbert space. 

Let us, then, judge the Barrett-Crane model in its natural classical setting, which is area Regge calculus. Note that in this setting, there is by definition no problem with the 4-simplex ``graviton propagator'': in area Regge calculus, there are no variables to vary in a 4-simplex other than its triangle areas. The question remains: on its own terms, is area Regge calculus a faithful discrete version of GR?

\section{The temptations of area Regge calculus} \label{sec:concept}

\subsection{Finite-region puzzles in quantum gravity} \label{sec:concept:puzzles}

The difficulties in quantum gravity are both conceptual and technical. The conceptual problems are greatly simplified when dealing with observables at the boundary of asymptotically flat or asymptotically AdS spacetime. Quantum gravity then reduces to finding an S-matrix, or all possible boundary correlators, respectively. Supergravity, perturbative string theory and AdS/CFT have made great progress in defining and calculating such observables, albeit never in fully realistic settings. 

The harder questions in quantum gravity refer to finite regions within the spacetime's bulk. What happens near the singularity of a black hole? What is the precise fate (operationally!) of locality, causality and unitarity? What is the probabilistic content of quantum mechanics when applied to finite-region geometries? One could try to argue, given the success of the descriptions on the spacetime's boundary, that these are red herrings: precise physics exists only at infinity, and that's that. However, the questions persist. Observers most likely can fall into black holes. From that moment on, at least classically, infinity is inaccessible and the singularity unavoidable. More universally, modern cosmology implies that \emph{all} observers live in a world with a positive cosmological constant, making spatial (or null) infinity inaccessible to \emph{anyone}. In such a world, there is no sense in which infinity can be approached asymptotically: the cosmological constant provides a finite large-distance cutoff.

Approaches to quantum gravity like LQG and spinfoams try to take seriously the issues concerning finite regions. However, they fall short. Hamiltonians may be defined, transition amplitudes may be calculated, but the operational probabilistic interpretation of the theory remains unclear. A key factor in this is the operational meaning of a Hilbert space and its relation to finite-time transition amplitudes. 

\begin{figure}%
\centering%
\includegraphics[scale=0.75]{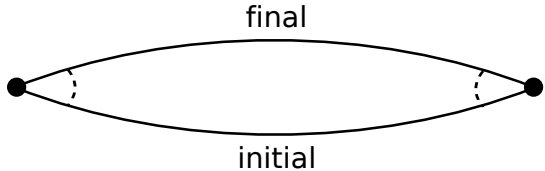} \\
\caption{A spacetime region with a spacelike closed boundary, composed of two hypersurfaces that intersect at a corner surface. When the geometry is dynamical, the only local measure of the ``time separation'' between the initial and final hypersurfaces is their intersection angle.}
\label{fig:spacelike_boundary} 
\end{figure}

When conducting experiments, one essentially prepares a state and then measures it at a later time. The probabilities of different measurements are then determined by quantum-mechanical amplitudes. The inner product in the Hilbert space of states \emph{at a given time} is obtained in the limit, when the preparation and measurement are (almost) simultaneous. In quantum field theory, finite-time transition amplitudes are usually associated with spacetime regions as in figure \ref{fig:spacelike_boundary}. The region's boundary consists of an initial 3d spacelike hypersurface and a final one. The two hypersurfaces intersect at a 2d surface. This corner surface is often taken at spacelike infinity, but this is not essential. The Hilbert-space product is obtained from the transition amplitude as the two hypersurfaces draw closer and closer together.

Now, in gravity, the relative positions of the initial and final hypersurfaces in spacetime cannot be easily specified a-priori. They are nonlocal functions of the metric in the bulk of the spacetime region, which is in turn dynamically determined by the initial and final data. Thus, without solving the dynamics, we cannot control the time separation between the two hypersurfaces! It's even impossible to control its sign: the bulk solution may be such that the two hypersurfaces intersect, thus flipping their ``initial'' and ``final'' roles. In particular, we can't easily construct the limit where the two hypersurfaces approach each other. As a result, it isn't clear how to give meaning to a Hilbert-space inner product.

A notion of time separation that \emph{does} persist in the gravitational case is the corner angle between the two hypersurfaces at their intersection. This is a local quantity, encoding a singularity in the boundary's extrinsic curvature. The corner angle can be cast as an actual time difference by passing to a Rindler frame in the infinitesimal neighborhood of the intersection surface. Of course, specifying the corner angle doesn't help with controlling the time separation throughout the boundary: away from the corner, the remarks from the previous paragraph still apply. The fact that time separation can be cleanly defined only near a 2d surface suggests that state spaces for finite-region gravity should be defined more naturally in terms of such surfaces. This is of course yet another intuitive argument for holography.

\subsection{How area Regge calculus fits in}

The metric discontinuities in area Regge calculus are usually seen as puzzling, at best. We will now point out two ways in which these discontinuities may help in addressing the issues raised in section \ref{sec:concept:puzzles}. These and other issues concerning quantum gravity in finite regions are the major conceptual open problem in theoretical physics. Area Regge calculus doesn't provide answers, but as we hope to show, it appears more suitable than standard Regge calculus as a starting point in the discussion.

\paragraph{The two geometries induced on a hypersurface.}

Recall from section \ref{sec:review:AreaRegge} that in area Regge calculus, a hypersurface can have two different geometries, induced by the 4-simplices on its two sides. If the hypersurface is spacelike (or null), these two geometries can be designated as ``initial'' and ``final''. We thus have an instantaneous evolution of the hypersurface geometry, where it's clear that the initial and final geometries are indeed on the same hypersurface in spacetime. As discussed in section \ref{sec:concept:puzzles}, it is this notion that appears necessary in order to define a Hilbert space of hypersurface geometries.

\paragraph{Angular ``time evolution'' of a triangle}

Let us now turn to the second notion of ``time separation'' discussed in section \ref{sec:concept:puzzles}: intersection angles between hypersurfaces. In Regge calculus, this translates into dihedral angles between tetrahedra within a triangle's normal plane (see figure \ref{fig:triangle}). In \emph{area} Regge calculus, as we rotate in the normal plane, the triangle's geometry undergoes shear discontinuities across 4-simplex boundaries. In other words, the triangle's conformal metric \emph{evolves} as a function of the angular (Rindler) time coordinate. Moreover, the equations of motion of area Regge calculus set the deficit angle around the triangle to zero. This means that the periodicity of the angular coordinate is always the same, with no dependence on the specific solution. Thus, around every triangle, there is a predetermined notion of Rindler time, with different solutions inducing different histories of the triangle's conformal metric.

\section{Nonzero scalar curvature: a failure of area Regge calculus} \label{sec:Ricci}

We now return to the central question: is area Regge calculus a discretization of GR? To address it, we must have some notion of Riemann curvature in a simplicial geometry with metric discontinuities. It was pointed out in \cite{Wainwright:2004yn} that the curvature of a discontinuous metric is ill-defined. Since the Christoffel symbols contain derivatives of the metric, they become distributional. The Riemann tensor must then contain products of these distributions, which are ill-defined in general.

One shouldn't conclude from this singular nature of the curvature that area Regge calculus cannot describe GR. Indeed, standard Regge calculus is singular in its own way. There, a closed loop encircling a triangle carries a deficit angle. This angle doesn't change as we shrink the loop to an arbitrarily small area. One could therefore say that the Riemann curvature diverges! As discussed in \cite{McDonald:2008js}, the Regge action avoids this problem by effectively always drawing the loop around a face of the Voronoi (dual circumcentric) lattice. In this way, the loop is kept at a natural, finite area. In the area calculus, a similar kind of smearing may be able to defuse the distribution-squared nature of the curvature at tetrahedra.   

Let us provisionally take this point of view. Also, as in standard Regge calculus, let us treat the curvature as if it is concentrated around the triangles. This seems more promising than looking at the discontinuity across each tetrahedron separately. A single triangle can be viewed as maintaining its 2d plane and its area, while its conformal metric evolves as a function of the angle in the normal plane. A triangle-based description successfully captures the discontinuity across each tetrahedron - once for each of the tetrahedron's faces. Moreover, it is at triangles that the on-shell vanishing of deficit angles takes effect. These considerations suggest that we focus on the metric around a single triangle. This metric can be written as: 
\begin{align}
 ds^2 = dr^2 + r^2 d\phi^2 + \gamma_{ij}(\phi) dx^i dx^j \ . \label{eq:metric}
\end{align}
Here, the $x^i$ are coordinates in the triangle's plane, while $(r,\phi)$ are polar coordinates in the normal plane. Due to the vanishing deficit angle, $\phi$ has the standard $2\pi$ periodicity. Different signatures can be incorporated by allowing complex $(r,\phi)$. If the normal plane is Euclidean, $\phi$ ranges in $[0,2\pi)\mod 2\pi$. If the normal plane is Lorentzian, $\phi$ ranges in $(-i\infty + n\pi/2, i\infty + n\pi/2)$, with different values $n=(0,1,2,3)\mod 4$ in the plane's four quadrants \cite{SorkinThesis}. $r$ ranges in $[0,\infty)$ when the radius is spacelike and in $[0,i\infty)$ when it is timelike.

If the discontinuities across 4-simplex boundaries are taken literally, then the triangle's metric $\gamma_{ij}(\phi)$ makes sudden jumps at values of $\phi$ where there is a tetrahedron (see figure \ref{fig:triangle}). Here, we will consider a regularization where the discontinuities are smeared out in $\phi$, so that the triangle's shear $\del_\phi\gamma_{ij} \equiv \sigma_{ij}$ under $\phi$ evolution is finite.

The condition that the triangle's area is independent of $\phi$ can be encoded as a tracelessness condition:
\begin{align}
 \gamma^{ij}\del_\phi\gamma_{ij} \equiv \gamma^{ij}\sigma_{ij} = 0 \ , \label{eq:shear}
\end{align}
where $\gamma^{ij}$ is the inverse of $\gamma_{ij}$. The Christoffel symbols and the Riemann tensor of the metric \eqref{eq:metric} have the following non-vanishing components:
\begin{align}
 &\Gamma^r_{\phi\phi} = -r\ ;\quad \Gamma^\phi_{r\phi} = \frac{1}{r}\ ;\quad
 \Gamma^\phi_{ij} = -\frac{1}{2r^2}\sigma_{ij}\ ;\quad \Gamma^i_{\phi j} = \frac{1}{2}\sigma^i_j\ ; \label{eq:Gamma} \\ 
 &R_{ri\phi j} = -\frac{1}{2r}\sigma_{ij}\ ; \quad
 R_{\phi i\phi j} = \frac{1}{2}\del_\phi\sigma_{ij} - \frac{1}{4}\sigma_{ik}\sigma^k_j\ ; \quad
 R_{ijkl} = \frac{1}{4r^2}(\sigma_{ik}\sigma_{jl} - \sigma_{il}\sigma_{jk}) \ ,
\end{align}
where indices on $\sigma_{ij}$ are raised with $\gamma^{ij}$. The Ricci scalar reads:
\begin{align}
 R = \frac{1}{r^2}\left(\gamma^{ij}\del_\phi\sigma_{ij} - \frac{3}{4}\sigma_{ij}\sigma^{ij} + \frac{1}{4}(\sigma_i^i)^2 \right)
   = \frac{1}{4r^2}\sigma_{ij}\sigma^{ij} \ . \label{eq:R}
\end{align}
In the last equality, we used the area-preserving condition \eqref{eq:shear}. As anticipated above, if we take the discontinuities at the 4-simplex boundaries literally, $\sigma_{ij}$ becomes distributional, making \eqref{eq:R} ill-defined. On the other hand, if we smooth out the discontinuities into a finite $\sigma_{ij}$, then only the origin of the normal plane is singular (note that $\gamma_{ij}$ is indeterminate there). The null rays in a Lorentzian normal plane may be also singular, if they carry a nonvanishing $\sigma_{ij}$. However, unless there is a null tetrahedron along one of these rays, it makes sense to keep $\sigma_{ij}\neq 0$ only in the relevant (spacelike or timelike) quadrants.  

A smeared version of the Ricci scalar in area Regge calculus may get contributions not only from the local value \eqref{eq:R}, but also from the holonomy around the origin in the triangle's normal plane. In fact, this is where curvature comes from in standard Regge calculus. It turns out, however, that in the area calculus, this extra contribution vanishes. Consider transporting a vector $v^\mu$ along a closed loop in the $\phi$ direction. Under parallel transport with the connection \eqref{eq:Gamma}, the components of $v^\mu$ evolve as:
\begin{align}
 \frac{dv^r}{d\phi} = -rv^\phi\ ;\quad \frac{dv^\phi}{d\phi} = \frac{1}{r}v^r\ ;\quad \frac{dv^i}{d\phi} = \frac{1}{2}\sigma^i_j v^j \ .
\end{align}
Due to the vanishing deficit angle, the overall change in $(v^r,v^\phi)$ along a closed loop vanishes. There may be a nonzero change in $v^i$, which will contribute to (a smeared version of) the Riemann component $R_{r\phi ij}$. However, this component makes no contribution to the Ricci tensor. We conclude that the mean Ricci scalar around the triangle is fully accounted for by the local value \eqref{eq:R}.
 
Now, for spacelike $\gamma_{ij}$, the square $\sigma_{ij}\sigma^{ij}$ is always non-negative. If in addition $r^2$ has the same sign wherever $\sigma_{ij}\neq 0$ (recall that $r$ is imaginary for timelike radii), then the Ricci scalar \eqref{eq:R} has the same sign everywhere. In the simplicial geometry, spacelike $\gamma_{ij}$ and a constant sign of $r^2$ at the (smeared) discontinuities correspond to spacelike triangles and to tetrahedra that are either all spacelike or all timelike. In such a case, we conclude that the Ricci scalar $R$ has the same sign around all the triangles, and thus has a non-vanishing mean value. The only exception is when $\sigma_{ij} = 0$ everywhere, i.e. when the metric is flat with no discontinuities. Nonzero $R$ of course contradicts the field equations of pure GR. We conclude that area Regge calculus, at least in the sector with all triangles spacelike and all tetrahedra of a fixed signature, is not a discretization of GR. The non-trivial solutions of the area calculus in this sector do not satisfy Einstein's equations.

\section{Discussion} \label{sec:discuss}

In this paper, we pointed out a serious problem with area Regge calculus: it is prone to develop nonzero Ricci curvature. We focused on the Ricci scalar in particular, since its mean value is easier to define than that of the full Ricci tensor. The theory's Euclidean sector appears completely ruled out, since all non-trivial solutions are characterized by a positive Ricci scalar. In the Lorentzian, the sector with spacelike triangles and either all-spacelike or all-timelike tetrahedra is similarly ruled out. These cases cover the setups that are usually considered in the spinfoam literature. We conclude that spinfoam models that reduce to area Regge calculus rather than standard Regge calculus are very probably non-viable. This rules out the Barrett-Crane model.   

The refractive-wave solution of \cite{Wainwright:2004yn} evades our argument, since the discontinuity there is along \emph{null} tetrahedra. One may be tempted to construct a version of area Regge calculus where all tetrahedra are null. Unfortunately, this is impossible for combinatoric reasons \cite{Diamonds}: 4-simplices with null tetrahedra cannot triangulate spacetime, due to a mismatch between the numbers of ``spacelike'' and ``timelike'' wedges between tetrahedra. In addition, null tetrahedra lead to infinite dihedral angles, making the Regge action ill-defined. 

It remains logically possible that some sector of Lorentzian area Regge calculus, featuring both spacelike and timelike tetrahedra, may correspond to discretized GR. We consider this unlikely. One may also hope that the non-vanishing Ricci scalar hints at GR with a cosmological constant. However, the model has no characteristic length scale, except perhaps the scale of the discretization. Thus, even if the Ricci scalar somehow develops a \emph{constant} mean value, as is necessary for GR with cosmological constant, then the cosmological radius will be at the same scale as the discretization - an unwanted outcome in itself. 

Another important possibility is that our negative result is an artifact of the intuitive regularization scheme that lead to the metric \eqref{eq:metric}. If someone comes up with a better way to approach the metric discontinuities, it would be most welcome. As laid out in section \ref{sec:concept}, the special properties of area Regge calculus sit well with the author's intuitions about quantum gravity. We conclude with the hope that these properties may be incorporated in a different framework, or that our negative result may be somehow overturned.

\section*{Acknowledgements}		

I am grateful to Norbert Bodendorfer, Aruna Kesavan and Carlo Rovelli for discussions. This work is supported in part by the NSF grant PHY-1205388 and the Eberly Research Funds of Penn State.


\begin{thebibliography} {99}

\bibitem{Regge:1961px} 
  T.~Regge,
  ``General Relativity Without Coordinates,''
  Nuovo Cim.\  {\bf 19}, 558 (1961).

\bibitem{Regge:2000wu} 
  T.~Regge and R.~M.~Williams,
  ``Discrete structures in gravity,''
  J.\ Math.\ Phys.\  {\bf 41}, 3964 (2000)
  [gr-qc/0012035].

\bibitem{Rovelli:2004tv} 
  C.~Rovelli,
  ``Quantum gravity,''
  Cambridge, UK: Univ. Pr. (2004) 455 p

\bibitem{Thiemann:2007zz} 
  T.~Thiemann,
  ``Modern canonical quantum general relativity,''
  Cambridge, UK: Cambridge Univ. Pr. (2007) 819 p
  [gr-qc/0110034].

\bibitem{Perez:2012wv} 
  A.~Perez,
  ``The Spin Foam Approach to Quantum Gravity,''
  Living Rev.\ Rel.\  {\bf 16}, 3 (2013)
  [arXiv:1205.2019 [gr-qc]].

\bibitem{EPRL}
  J.~Engle, E.~Livine, R.~Pereira and C.~Rovelli,
  ``LQG vertex with finite Immirzi parameter,''
  Nucl.\ Phys.\  B {\bf 799}, 136 (2008)
  [arXiv:0711.0146 [gr-qc]].

\bibitem{FK}
  L.~Freidel, K.~Krasnov,
  ``A New Spin Foam Model for 4d Gravity,''
  Class.\ Quant.\ Grav.\  {\bf 25}, 125018 (2008).
  [arXiv:0708.1595 [gr-qc]].

\bibitem{Bianchi:2010gc} 
  E.~Bianchi, P.~Dona and S.~Speziale,
  ``Polyhedra in loop quantum gravity,''
  Phys.\ Rev.\ D {\bf 83}, 044035 (2011)
  [arXiv:1009.3402 [gr-qc]].

\bibitem{Rovelli:1993kc} 
  C.~Rovelli,
  ``The Basis of the Ponzano-Regge-Turaev-Viro-Ooguri quantum gravity model in the loop representation basis,''
  Phys.\ Rev.\ D {\bf 48}, 2702 (1993)
  [hep-th/9304164].

\bibitem{Barrett:1997tx} 
  J.~W.~Barrett, M.~Rocek and R.~M.~Williams,
  ``A Note on area variables in Regge calculus,''
  Class.\ Quant.\ Grav.\  {\bf 16}, 1373 (1999)
  [gr-qc/9710056].

\bibitem{Barrett:1994nn} 
  J.~W.~Barrett,
  ``First order Regge calculus,''
  Class.\ Quant.\ Grav.\  {\bf 11}, 2723 (1994)
  [hep-th/9404124].

\bibitem{Dittrich:2008va} 
  B.~Dittrich and S.~Speziale,
  ``Area-angle variables for general relativity,''
  New J.\ Phys.\  {\bf 10}, 083006 (2008)
  [arXiv:0802.0864 [gr-qc]].

\bibitem{Barrett:1997gw} 
  J.~W.~Barrett and L.~Crane,
  ``Relativistic spin networks and quantum gravity,''
  J.\ Math.\ Phys.\  {\bf 39}, 3296 (1998)
  [gr-qc/9709028].

\bibitem{Barrett:1999qw} 
  J.~W.~Barrett and L.~Crane,
  ``A Lorentzian signature model for quantum general relativity,''
  Class.\ Quant.\ Grav.\  {\bf 17}, 3101 (2000)
  [gr-qc/9904025].

\bibitem{Wainwright:2004yn} 
  C.~Wainwright and R.~M.~Williams,
  ``Area Regge calculus and discontinuous metrics,''
  Class.\ Quant.\ Grav.\  {\bf 21}, 4865 (2004)
  [gr-qc/0405031].

\bibitem{Barrett:2000jy} 
  J.~W.~Barrett,
  ``Refractive gravitational waves and quantum fluctuations,''
  gr-qc/0011051.

\bibitem{Baez:2002rx} 
  J.~C.~Baez, J.~D.~Christensen and G.~Egan,
  ``Asymptotics of 10j symbols,''
  Class.\ Quant.\ Grav.\  {\bf 19}, 6489 (2002)
  [gr-qc/0208010].

\bibitem{Freidel:2002mj} 
  L.~Freidel and D.~Louapre,
  ``Asymptotics of 6j and 10j symbols,''
  Class.\ Quant.\ Grav.\  {\bf 20}, 1267 (2003)
  [hep-th/0209134].

\bibitem{Barrett:2002ur} 
  J.~WBarrett and C.~M.~Steele,
  ``Asymptotics of relativistic spin networks,''
  Class.\ Quant.\ Grav.\  {\bf 20}, 1341 (2003)
  [gr-qc/0209023].

\bibitem{Bianchi:2006uf} 
  E.~Bianchi, L.~Modesto, C.~Rovelli and S.~Speziale,
  ``Graviton propagator in loop quantum gravity,''
  Class.\ Quant.\ Grav.\  {\bf 23}, 6989 (2006)
  [gr-qc/0604044].

\bibitem{Rovelli:2012yy} 
  C.~Rovelli and E.~Wilson-Ewing,
  ``Discrete Symmetries in Covariant LQG,''
  Phys.\ Rev.\ D {\bf 86}, 064002 (2012)
  [arXiv:1205.0733 [gr-qc]].

\bibitem{Christodoulou:2012sm} 
  M.~Christodoulou, A.~Riello and C.~Rovelli,
  ``How to detect an anti-spacetime,''
  Int.\ J.\ Mod.\ Phys.\ D {\bf 21}, 1242014 (2012)
  [arXiv:1206.3903 [gr-qc]].

\bibitem{Livine:2002rh} 
  E.~R.~Livine and D.~Oriti,
  ``Implementing causality in the spin foam quantum geometry,''
  Nucl.\ Phys.\ B {\bf 663}, 231 (2003)
  [gr-qc/0210064].

\bibitem{Barrett:2009mw} 
  J.~W.~Barrett, R.~J.~Dowdall, W.~J.~Fairbairn, F.~Hellmann and R.~Pereira,
  ``Lorentzian spin foam amplitudes: Graphical calculus and asymptotics,''
  Class.\ Quant.\ Grav.\  {\bf 27}, 165009 (2010)
  [arXiv:0907.2440 [gr-qc]].

\bibitem{Han:2013gna} 
  M.~Han and T.~Krajewski,
  ``Path Integral Representation of Lorentzian Spinfoam Model, Asymptotics, and Simplicial Geometries,''
  arXiv:1304.5626 [gr-qc].

\bibitem{Han:2013hna} 
  M.~Han,
  ``On Spinfoam Model in Large Spin Regime,''
  arXiv:1304.5627 [gr-qc].

\bibitem{Mikovic:2011zx} 
  A.~Mikovic and M.~Vojinovic,
  ``Effective action and semiclassical limit of spin foam models,''
  Class.\ Quant.\ Grav.\  {\bf 28}, 225004 (2011)
  [arXiv:1104.1384 [gr-qc]].

\bibitem{Engle:2011un} 
  J.~Engle,
  ``A proposed proper EPRL vertex amplitude,''
  arXiv:1111.2865 [gr-qc].

\bibitem{Alesci:2007tx} 
  E.~Alesci and C.~Rovelli,
  ``The Complete LQG propagator. I. Difficulties with the Barrett-Crane vertex,''
  Phys.\ Rev.\ D {\bf 76}, 104012 (2007)
  [arXiv:0708.0883 [gr-qc]].

\bibitem{McDonald:2008js} 
  J.~R.~McDonald and W.~A.~Miller,
  ``A Discrete Representation of Einstein's Geometric Theory of Gravitation: The Fundamental Role of Dual Tessellations in Regge Calculus,''
  arXiv:0804.0279 [gr-qc].

\bibitem{SorkinThesis}
  R.~Sorkin,
  ``Development of simplectic methods for the metrical and electromagnetic fields,''
  Ph.D. thesis, California Institute of Technology, 1974. 

\bibitem{Diamonds} 
  Y.~Neiman,
  ``Causal cells: spacetime polytopes with null hyperfaces,''
  Geometriae Dedicata, 0046-5755 (2013), 
  DOI:10.1007/s10711-012-9823-0
  [arXiv:1212.2916 [math.CO]].

\end{thebibliography}
\end{document}